\documentclass[11pt,preprint]{aastex6}
\defcitealias{2013ApJ...777L...5C}{C13}

\shorttitle{Local Disk Velocity Substructure}
\shortauthors{Pearl, Newberg, Carlin}

\begin{document}
\title{A Map of the Local Velocity Substructure in the Milky Way Disk}

\author{
	Alan N. Pearl\altaffilmark{1}, 
	Heidi Jo Newberg\altaffilmark{1}, 
	Jeffrey L. Carlin\altaffilmark{2}, 
	\& R. Fiona Smith\altaffilmark{1}
}
\altaffiltext{1}{Department of Physics, Applied Physics and Astronomy,
Rensselaer Polytechnic Institute, Troy, NY 12180, USA}
\altaffiltext{2}{LSST, 933 North Cherry Avenue, Tucson, AZ 85721, USA}


\begin{abstract}

We confirm, quantify, and provide a table of the coherent velocity substructure of the Milky Way disk within 2~kpc of the Sun towards the Galactic anticenter, with 0.2 kpc resolution.
We use the radial velocities of $\sim$340,000 F-type stars obtained with the Guoshoujing Telescope (also known as the Large Sky Area Multi-Object Fiber Spectroscopic Telescope, LAMOST), and proper motions derived from the PPMXL catalog.
The PPMXL proper motions have been corrected to remove systematic errors by subtracting the average proper motions of galaxies and QSOs that have been confirmed in the LAMOST spectroscopic survey, and that are within 2.5$^{\circ}$ of the star's position.
We provide the resulting table of systematic offsets derived from the PPMXL proper motion measurements of extragalactic objects identified in the LAMOST spectroscopic survey.
Using the corrected phase-space stellar sample, we find statistically significant deviations in the bulk disk velocity of 20~km~s$^{-1}$ or more in the three dimensional velocities of Galactic disk stars.  The bulk velocity varies significantly over length scales of half a kpc or less.  The rotation velocity of the disk increases by 20~km~s$^{-1}$ from the Sun's position to 1.5 kpc outside the solar circle.  Disk stars in the second quadrant, within 1~kpc of the Sun, are moving radially towards the Galactic center and vertically towards a point a few tenths of a kpc above the Galactic plane; looking down on the disk, the stars appear to move in a circular streaming motion with a radius of order 1 kpc.

\end{abstract}

\section{Introduction}
\subsection{Disk Substructure}

Recent results have shown that the stellar disk of the Milky Way exhibits coherent substructure in both spatial density and velocity.  Using data from the Sloan Digitial Sky Survey \citep[SDSS;][]{2000AJ....120.1579Y} and the companion Sloan Extension for Galactic Understanding and Exploration \citep[SEGUE;][]{2009AJ....137.4377Y}, both \citet{2012ApJ...750L..41W} and \citet{2013ApJ...777...91Y} found that at the position of the Sun there are more stars 1~kpc above the Galactic plane than 1~kpc below it, but fewer stars 400~pc above the Galactic plane than 400~pc below it.  This points to a ``bending'' mode of the disk (``breathing" modes are symmetric in density about the Galactic plane).  SDSS/SEGUE also uncovered vertical velocity substructure \citep{2012ApJ...750L..41W}, and oscillations of the stellar phase-space density \citep{2012MNRAS.423.3727G} that were thought to be induced by minor merger events in the Milky Way.  Data from the RAdial Velocity Experiment \citep[RAVE;][]{2006AJ....132.1645S} allowed \citet{2013MNRAS.436..101W} to discover that the disk exhibits wave-like substructure in the $V_Z$, $V_R$, and $V_\theta$ directions, with amplitudes as high as 17 km s$^{-1}$.  These disk substructures are also present in data from the Chinese Large Area Multi-Object Spectroscopic Telescope (LAMOST, also named the Guo Shou Jing Telescope after a Chinese astronomer and engineer who is famous for his instrumentation; \citealt{2012RAA....12.1197C}; \citealt{2004ChJAA...4....1S}).  \citet{2013ApJ...777L...5C} used LAMOST radial velocities to show that in the vicinity of the Sun the stars above the Galactic plane exhibited net flow outward from the Galactic center and downward towards the Galactic plane.  Stars below the disk showed bulk flow in the opposite direction.  Although all of the surveys showed similar types of structure, they used different tracers and probed different portions of the Galactic disk. It has been difficult to combine the results of each publication into a global view of disk substructure.

\citet{2015ApJ...801..105X} found that the midplane of the disk oscillates vertically, as one looks towards the Galactic anticenter from the position of the Sun, indicative of a bending mode of the disk.  This was later supported by \citet{2016ApJ...825..140M}, using data from the Panoramic Survey Telescope and Rapid Response System 1 Survey \citep[PS1;][]{2010SPIE.7733E..0EK} $3\pi$ dataset.  These ripples in the disk roughly follow the pattern of our Galaxy's spiral arms, as traced by HII regions, giant molecular clouds, and methanol masers \citep{2014A&A...569A.125H}.  The Monoceros Ring (\citealt{2003ApJ...588..824Y}, \citealt{2003MNRAS.340L..21I}) at 20~kpc from the Galactic center, and the more distant TriAnd Cloud \citep{2004ApJ...615..732R} might also be explained as ripples in a stellar disk that extends at least 25~kpc from the Galactic center.  This latter suggestion is supported by the observation that the TriAnd and TriAnd2 substructures have metallicities indicative of association with a disk population rather than a dwarf satellite \citep{2015MNRAS.452..676P}.

Although there might be other mechanisms for creating density and velocity substructure in the disk \citep{2014MNRAS.443L...1D, 2014MNRAS.440.2564F, 2015MNRAS.452..747M, 2016MNRAS.461.3835M}, the observed oscillations are roughly consistent with simulations of the disk's response to a large ($>10^9 M_\odot$) satellite or dark subhalo passing near, or through, the disk (\citealt{2008ApJ...688..254K}, \citealt{2008ApJ...676L..21Y}, \citealt{2009MNRAS.396L..56M}, \citealt{2011Natur.477..301P}, \citealt{2013MNRAS.429..159G}, \citealt{2014MNRAS.440.1971W}, \citealt{2016ApJ...823....4D}, \citealt{2016arXiv160804743L}).  Corrugations have also been observed in HI gas and young stellar populations in the Milky Way \citep{1986A&A...163...43S,2006ApJ...643..881L}, and in external galaxies \citep{2008AJ....135..291M,2008ApJ...688..237M}.  It has been proposed that these oscillations were induced by dwarf galaxy satellites \citep{2011ApJ...743...35C}.  There is the potential that disk substructure could be used to probe the properties of Milky Way satellites, even if the satellites are composed of dark matter and are not directly visible \citep{2015MNRAS.446.1000F}, a possibility called {\it galactoseismology} by \citet{2012ApJ...750L..41W}.  \citet{2015ApJ...802L...4C} used observed ripples in the Milky Way gas to infer the presence of a previously undetected dwarf satellite of the Milky Way, though the existence of this dwarf galaxy is disputed \citep{2015ApJ...813L..40P}. 

\citet{2016MNRAS.456.2779G, 2017MNRAS.465.3446G} examined fully cosmological simulations of Milky Way-like galaxies, and conclude that satellite-induced vertical oscillations of the disk are common.  In addition, they can explain density substructures like the ones observed in the Monoceros Ring.  More impressively, they show that as a satellite falls into the Milky Way, it can perturb the density substructure of the halo, and that can in turn produce perturbations in the disk (see also \citealt{1998MNRAS.299..499W, 2000ApJ...534..598V}).  This means that previous simulations of the effects of the Sagittarius dwarf galaxy infall on the disk were missing an important contribution -- the effect of a redistribution of halo mass on the disk.  This might explain why previous simulations had difficulty producing the magnitude of the disk oscillations observed in the Milky Way.

The similarity between the observed ripple pattern, the Milky Way's spiral arms, and the spiral structure produced in the simulations, leads one to wonder whether the spiral structure in galaxies is induced and sustained by orbiting dwarf galaxies (and possibly dark subhalos).  Although it has been known for some time that galaxy interactions can produce spiral arms \citep{1972ApJ...178..623T}, the source of spiral structure like that of the Milky Way has traditionally been explained under the assumption of quasi-stationary spiral structure (QSSS; \citealt{2016ARA&A..54..667S}).  The possibility that dwarf galaxies could be exciting the disk substructure observed in the Milky Way, and the knowledge that the dwarf galaxy interactions also produce spiral structure, could mean that the QSSS hypothesis, and thus spiral density waves, may not be required to sustain spiral structure in Milky Way-like galaxies \citep{2017IAUS..321...13N}.  In support of this, note that \citet{2015MNRAS.454..933D} show that disk substructure generated from dwarf galaxy interactions is the result of the phase wrapping of epicyclic perturbations, without the self-gravity required by wave propagation in breathing or bending modes.
While some effort has gone into understanding the effects of the Galactic bar and existing spiral arms on disk substructure (e.g. -- \citealt{2014MNRAS.440.2564F, 2014MNRAS.443L...1D, 2016MNRAS.461.3835M, 2016ApJ...824...39V}), the connection between satellites and the origin of normal spiral structure has not been explored.

In this paper, we reproduce the velocity substructure within $\sim 1.5$~kpc of the Sun, following a technique and data set very similar to that of \citet{2013ApJ...777L...5C}.  However, we have more data, greatly improved distance estimates, improved proper motions, and we propagate errors in our measurements to produce a map of the velocity substructure in a portion of the local disk.  Our maps confirm the compression of the disk reported outside of the solar circle by \citet{2013MNRAS.436..101W}, though we observe the compression in the third quadrant and not in the second.  The maps also confirm the nearer bulk motion of the disk found by \citet{2013ApJ...777L...5C}; the more distant parts of the Carlin~et~al. data had larger velocity errors and were not used in our study.  We observe an increase in disk rotation speed at greater radii outside the solar circle, which is comparable to the trend found in \citet{2014A&A...563A.128L}, but we map the trend in three dimensions.

We present a three-dimensional map, with errors, of the disk velocity substructure in a region approximately $1.5 \times 1.5 \times 1.5$~kpc in size, centered on the Galactic plane and extending out from the Sun towards the anticenter.  While the disk appears to have zero Galactocentric radial velocity in the third quadrant, the stars in the second quadrant, just half a kpc away, exhibit a bulk motion towards the Galactic center of about 15~km~s$^{-1}$ and also a compression in the $Z$~direction towards a point about 0.5~kpc above the Galactic plane.  We observe that the magnitude of the circulation velocity decreases as a function of distance from the Galactic plane, as has been known for many years.  The circulation velocity of stars within 0.2 kpc of the Galactic plane is 205 km s$^{-1}$ near the Sun's position, and increases to 225 km s$^{-1}$ 1.5 kpc farther from the Galactic center.  The velocity substructure map we present is the best to date because it shows the statistical significance of the measured bulk velocity with a granularity of 0.2~kpc in a local small region.

\subsection{The Large Area Multi-Object Spectroscopic Telescope (LAMOST)}

LAMOST \citep{2012RAA....12.1197C} is one of the National Major Scientific Projects undertaken by the Chinese Academy of Sciences. It is a 4~meter class, downward-facing Schmidt reflector with 4000~optical fibers in the focal plane, that was designed for spectroscopic surveys.  The spectra have resolution $R = \lambda/\Delta\lambda = 1800$, which is similar to the resolution of SDSS/SEGUE. The telescope is located at the Xinglong Observing Station ($40^\circ23'39''$N $117^\circ34'30''$E), one hundred miles northeast of Beijing.  Due to the fixed second mirror (called ``mirror B'' because it is the second reflecting surface but corresponds to the primary mirror of a Schmidt telescope), the optics favor targets closer to the zenith.  Local weather limits the observing season to the months September to May. As a result of the fixed mirror~B and restricted observing season, survey observations are restricted to declinations $-10^\circ<\delta<60^\circ$, and right ascensions near the Galactic anticenter are preferred. 

The optical properties of the telescope system and the site location limit the magnitudes of spectroscopic targets to $V \sim 19.5$ \citep{2012RAA....12.1243L}, though most of the observed objects are close to $V \sim 16$.  The telescope has been used to carry out a spectroscopic survey \citep{2012RAA....12..723Z} that primarily yields Milky Way stars, and most of those stars are in the Galactic disk.  For example, in the 2014/2015 observing season, the survey netted 1,558,924~spectra, of which 1.4M are classified as Galactic stars, 23K are classified as galaxies, and 7K are classified as QSOs \citep{2016arXiv160102334H}. Details of the LAMOST spectral survey can be found in \citet{2012RAA....12..723Z}.  The science plan for the LEGUE portion of the survey can be found in \citet{2012RAA....12..735D}. The pilot survey began in October 2011, and the general survey started in September 2012 and is still in progress.  The survey has publicly released the pilot survey \citep{2012RAA....12.1243L}, Data Release 1 (DR1; \citealt{2015RAA....15.1095L}), and Data Release 2 (DR2; \citealt{2016yCat.5149....0L}).  LAMOST DR2 includes measured parameters for 3.9M spectra obtained through spring 2014.

The data used in this work include stellar parameters measured by the LAMOST pipeline for all stars in LAMOST Data Release 3 (DR3, which includes data from the first 3 years of survey operations plus the Pilot Survey), and data from the first 2 quarters of the fourth year of survey operations (data observed through Feb 2016). This encompasses a total of $\sim3.9$~million stellar spectra of $\sim3.1$~million unique stars. Extinction corrections throughout are based on the maps of \citet{1998ApJ...500..525S}. In addition to the measured stellar parameters, distances derived via the method of \citet{2015AJ....150....4C} are included in our analysis. For stars that have been observed multiple times, the stellar parameters and distances represent weighted mean values of all measurements, weighted by spectroscopic $S/N$. 

\section{Systematic Proper Motion Correction}
\subsection{Systematics of PPMXL Proper Motions}

Positions and Proper Motions ``Extra Large'' \citep[PPMXL;][]{2010AJ....139.2440R} is a full-sky catalog containing the positions and proper motions for about 900~million objects.
However, it is known (\citealt{2011PASP..123.1313W}, \citealt{2015RAA....15..849G}, \citealt{2016AJ....151...99V}) that the proper motions calculated by PPMXL contain significant systematic error.
We again confirm this systematic error by examining the PPMXL proper motion measurements of quasars and galaxies identified by LAMOST.  Since these objects are far too distant to contain any true proper motion, the proper motion measurements result entirely from a combination of statistical and systematic error.

Any systematic error in proper motion measurements within a particular region of the sky causes a systematic shift in the perpendicular component of velocity that increases linearly with distance.
This can become very significant, even for observations within the local disk region (1~kpc~$\times$~1~mas~yr$^{-1} = 4.74$~km~s$^{-1}$).
Therefore, it is extremely important to remove as much of this systematic error as possible before calculating the three dimensional stellar velocities.

Published proper motion corrections are based on a variety of fitting methods of extragalactic objects with varying ranges of precision, from constant shifts \citep{2011PASP..123.1313W} to correction tables \citep{2015RAA....15..849G}.
Other corrections implement analytic fitting methods (\citealt{2013ApJ...777L...5C}, \citealt{2016AJ....151...99V}), which are effective in smoothing out large scale systematics, but do not necessarily work well in all areas of the sky; especially in low Galactic latitude regions.
In this work, we analyze the correction from \citet{2016AJ....151...99V}, which is magnitude dependent and utilizes spherical harmonics fits, since it is the previous best global correction for the PPMXL database.

\subsection{Extragalactic Sources from LAMOST} \label{sec:qso_selection}

\begin{figure}
\centering
\includegraphics[width=.7\textwidth]{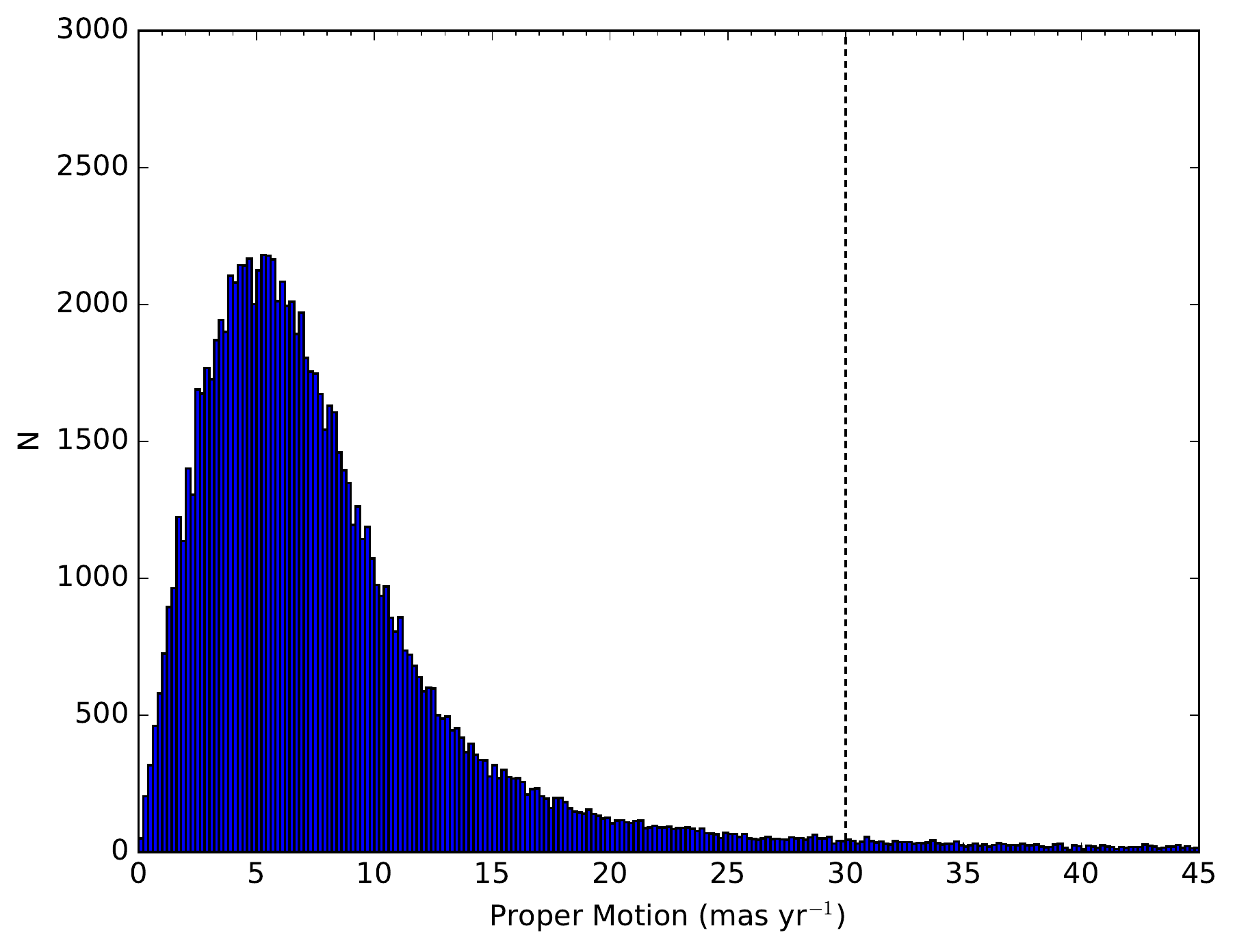}
\caption{
	Distribution of proper motion magnitudes of extragalactic objects as calculated by PPMXL.
	Objects with proper motions greater than the vertical grid line at $|\mu|=30$ were not used in our correction due to the greater likelihood that they were misidentified.
	Additionally, this removes outliers from our data set, to which our correction may be very sensitive.
	\label{fig:pm_hist}
}
\end{figure}

The template matching algorithm in the LAMOST pipeline categorizes objects based on their best template match as ``STAR,'' ``GALAXY,'' ``QSO,'' or ``UNKNOWN,'' where the UNKNOWN category is a catch-all for spectra that do not match any of the templates well. To derive proper motion corrections, we select all extragalactic objects classified as GALAXY or QSO in the data set up through DR4Q2. These should constitute a fixed reference frame for proper motion zero points. As with the stellar catalog, we match the extragalactic objects to PPMXL with a $1.5''$ matching radius.
We remove all duplicates and select objects with a proper motion less than $|\mu|_{\rm max} = 30$~mas~yr$^{-1}$ (Figure~\ref{fig:pm_hist}; the determination of this value is explained further in Section~\ref{sec:method}) in order to remove outliers and possibly nearby stars with large proper motion that were misidentified.
After all selections, we obtained a sample of 100,367~LAMOST spectra of quasars and galaxies, along with their corresponding proper motions from the PPMXL catalog.

\begin{figure}
\centering
\includegraphics[width=.8\textwidth]{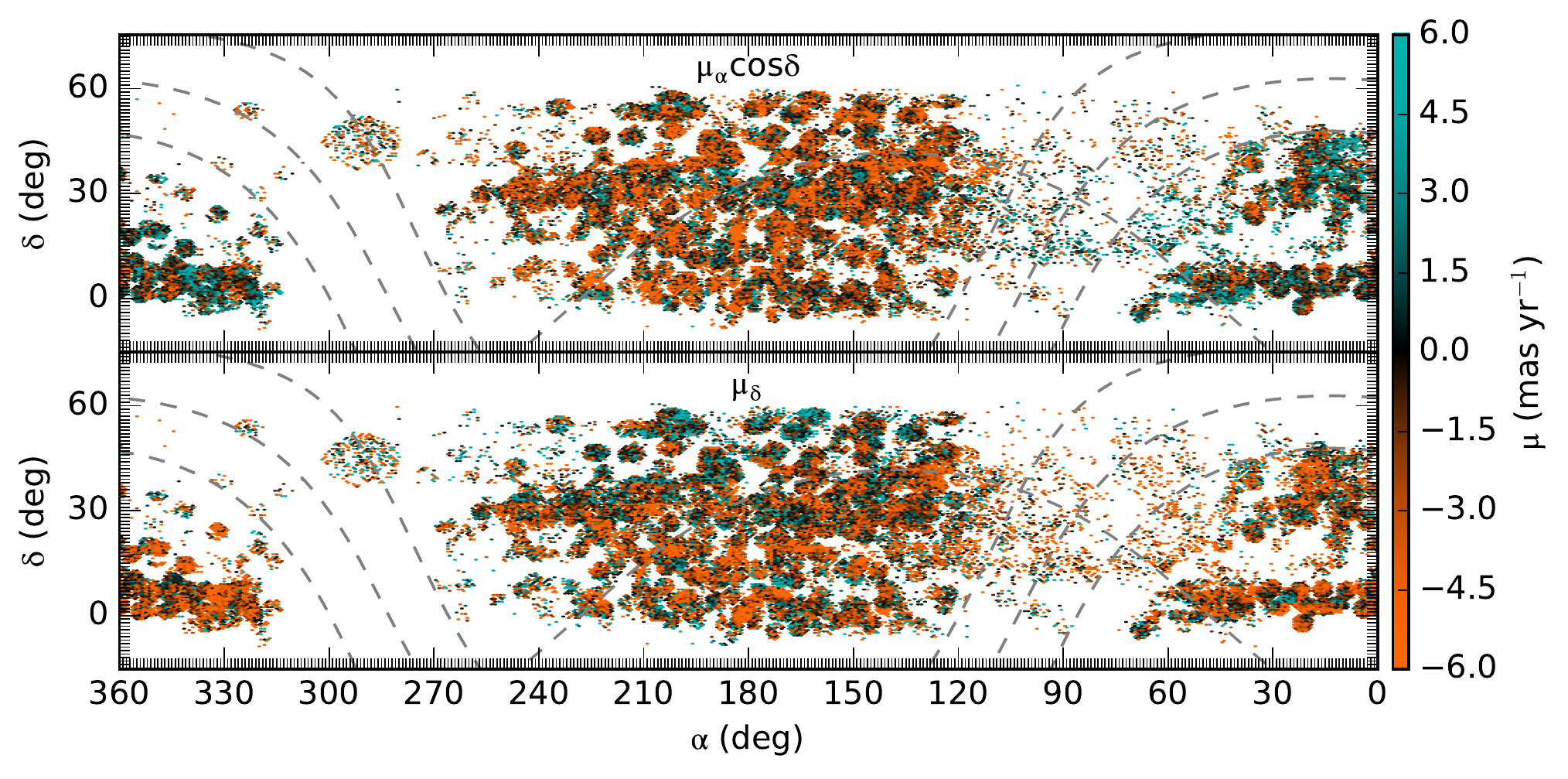}
\caption{
	Measured proper motion of extragalactic objects by sky coordinates.
	Notice many randomly scattered areas on the order of several degrees in which there is clumping of blue or orange, indicating a systematic error in that region of the sky.
	The top and bottom panels show proper motion in the right ascension and declination components, respectively.
	The grey grid lines represent the galactic latitude lines of zero and $\pm15^{\circ}$, as well as the longitude line of zero and 180$^{\circ}$, with the galactic anticenter at the intersection point and the northern galactic pole in the center of the plot.
	\label{fig:qag_none}
}
\end{figure}
\begin{figure}
\centering
\includegraphics[width=.8\textwidth]{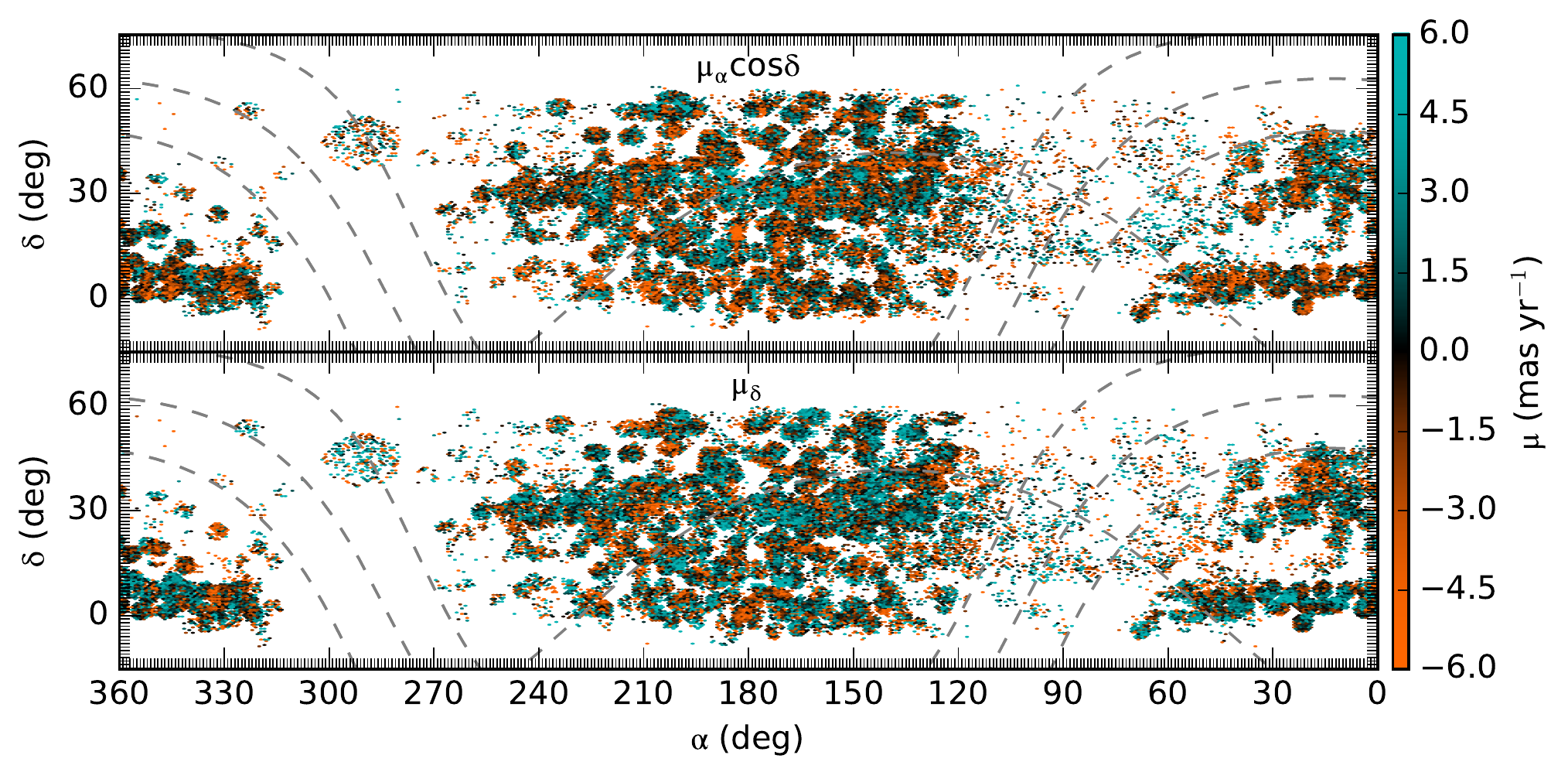}
\caption{
	Corrected \citep{2016AJ....151...99V} proper motion of extragalactic objects by sky coordinates.
	Overall, this shows a significant improvement over Figure~\ref{fig:qag_none}, however, there are still many areas with very noticeable systematic error, mostly due to overcorrection.
	This occurs very often in regions which originally had very little error, but were nearby to regions with high error.
	\label{fig:qag_vick}
}
\end{figure}

Since the true proper motion of these objects must be zero, we expect to see an approximately uniform Gaussian random distribution around zero along each direction to account for statistical error.
From a quick glance at a plot of proper motions over the sky (Figure~\ref{fig:qag_none}), it is very clear that there is systematic clumping of proper motions based on their celestial coordinates, in addition to an overall negative net motion in each component.
In comparison, the \citet{2016AJ....151...99V} correction significantly decreases the error in most parts of the sky.
However, we show that there are still many parts of the sky where this error is actually increased (Figure~\ref{fig:qag_vick}).
This occurs very often in the surrounding area of regions with large systematic error, indicating that the systematics of these objects deviate over a scale too small to be handled well by an analytic function.


\subsection{PPMXL Proper Motion Correction Method} \label{sec:method}

Due to the apparently random small angle variations in the systematic error (see Section~\ref{sec:corr_analysis}), fitting the proper motions of extragalactic objects to any kind of analytic function would need to be of extremely high order to ensure an improvement in all regions of the sky.
It is also unclear, using this method, which regions of the sky can be confidently used without a function to give an estimate of remaining systematic error.
Giving a quantifiable estimation for this remaining systematic error is absolutely necessary in measuring the statistical significance of stellar velocities.
An alternative to this method, which we choose to use in this work, is to construct a table of systematic offsets over many bins across the sky, independently calculating the correction needed at each position to produce a result somewhat similar to \citet{2015RAA....15..849G}.

The concept of our correction is to construct a large table of corrections over right ascension $\alpha$ and declination $\delta$, where at each correction point ($\alpha$,$\delta$), we return correction values based on the systematics of extragalactic objects near this position, but not necessarily only those within the region constrained by the bin.
From a given correction point, we search within a surrounding search radius for nearby quasars and galaxies from our LAMOST data.
The opposite of the mean of the proper motions of these objects is calculated as the systematic correction value for this point, in each component.
This is the value that we add to any object in the 0.25$^{\circ}$ by 0.25$^{\circ}$ bin centered at this coordinate (see Section~\ref{sec:corr_table}).

Using this method, there are a few subjective parameters which we define here:
Since our method is fairly vulnerable to outliers and the LAMOST data set contains objects with proper motion upwards of several hundred~mas~yr$^{-1}$, we initially eliminate all objects with a proper motion above an upper bound of $|\mu|_{\rm max} = 30$~mas~yr$^{-1}$ during our data selection (Section~\ref{sec:qso_selection}).
We also define a search radius $r_{\rm search} = 2.5^{\circ}$ (Figure~\ref{fig:dist_vs_match}), which is the maximum angular separation allowed in order to be considered a nearby object and used in the correction.
Objects further apart than this are not used in the correction.
In order to eliminate any rare instances of two or three objects with large and similar random error mimicking a precise and accurate correction, we set a minimum number of objects, $n_{\rm min} = 4$, that must be found in order to return a correction.
Lastly, in order to cut down on excessive computation time, we stop searching after the nearest $n_{\rm max} = 100$ objects are found.


We tested many different values for $|\mu|_{\rm max}$ and $r_{\rm search}$ before arriving at these values in order to optimize the correction. 
They were tested by attempting to correct each object using only its nearby objects, and not itself, to ensure self correction did not contribute to the success of the entire correction.
For testing purposes, the correction points were defined to be the positions of each extragalactic object with which we tested the algorithm; for the correction table, correction points were selected to be at the centers of each $0.25^\circ \times 0.25^\circ$ bin.
Using the resulting correction, we optimized two main criteria:
Firstly, a lower value for the average magnitude of proper motion would indicate high precision in the values output by the correction; however, most of this is due to random error which is impossible to correct.
Secondly and more importantly, a normal distribution of positive and negative values centered around zero in each proper motion component, independent of position, would indicate a lack of systematic error.

\begin{figure}
\centering
\includegraphics[width=.8\textwidth]{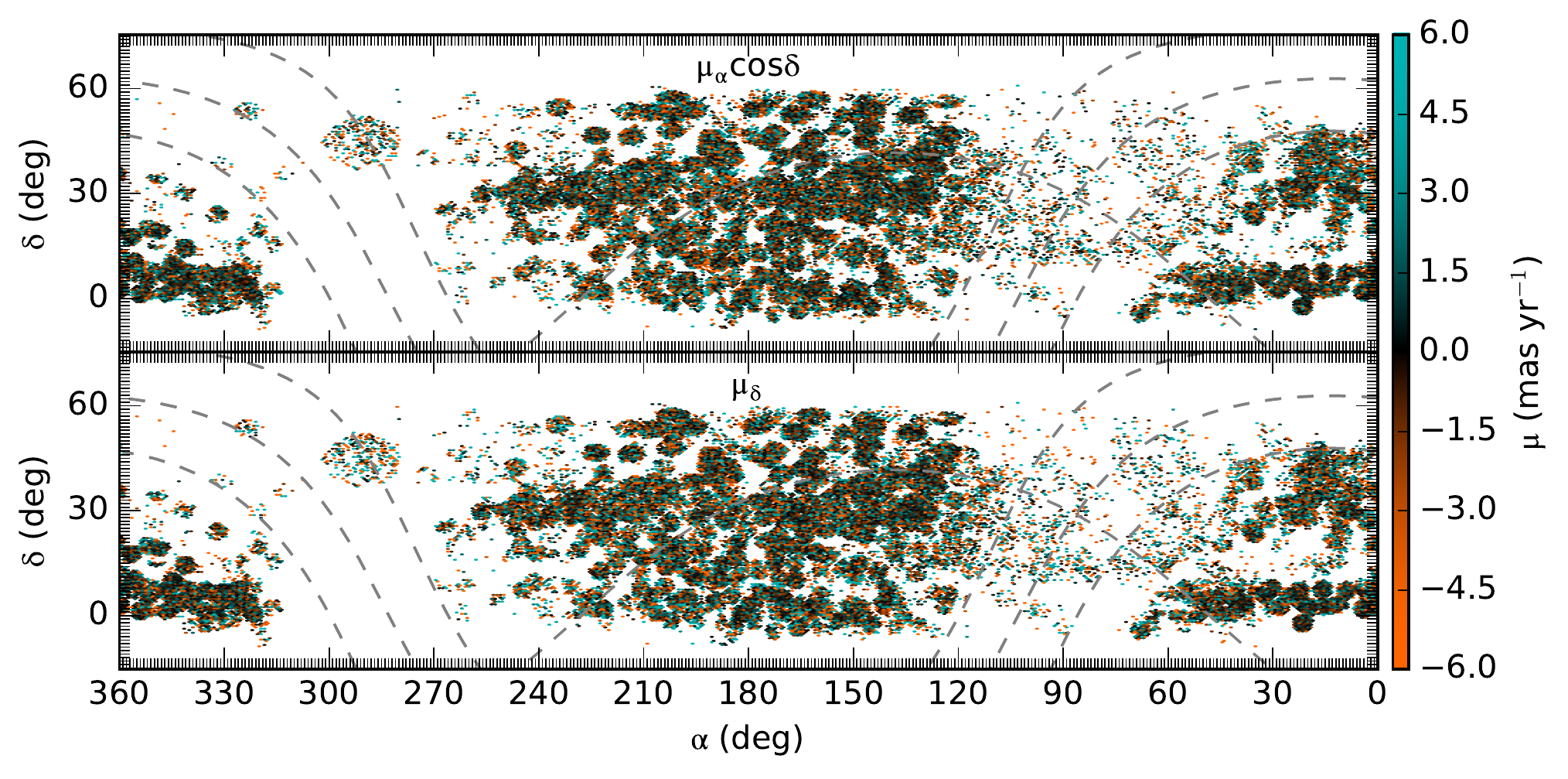}
\caption{
	Corrected (using this work) proper motion of extragalactic objects by sky coordinates.
	The distribution of these values resemble random error around zero significantly more than those in Figure~\ref{fig:qag_none} in all observed regions of the sky.
	Each object in this plot was corrected using our correction method described in Section~\ref{sec:method}, but without factoring in its own proper motion to ensure self correction was not the cause of any improvement.
	\label{fig:qag_pearl}
}
\end{figure}

\begin{figure}
\centering
\includegraphics[width=.7\textwidth]{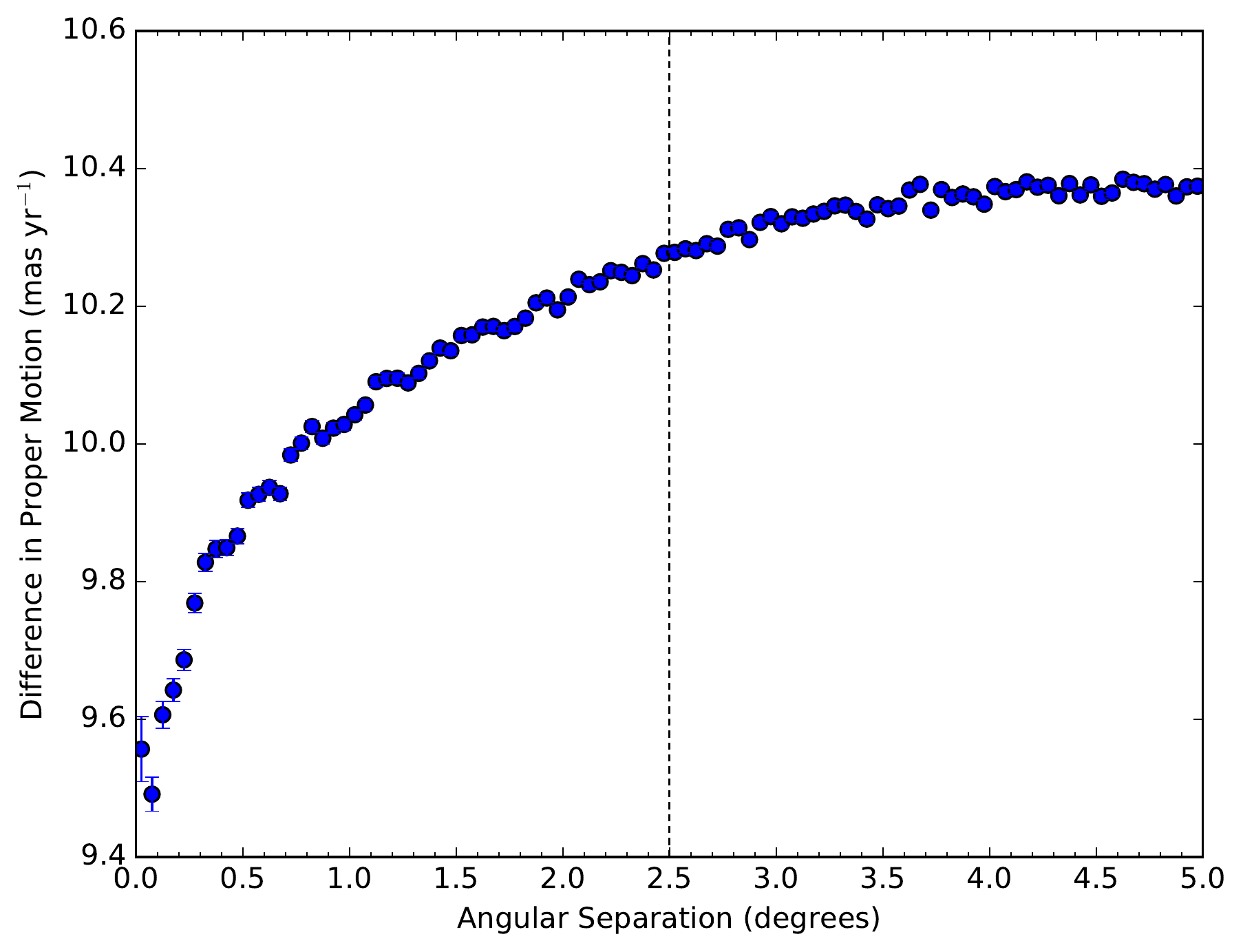}
\caption{
	Magnitude of difference in proper motion vs. angular separation between extragalactic objects.
	Random sample of proper motion comparisons between every pair of quasars, binned every 0.05~degrees of angular separation.
	This shows that objects closer together have a higher correlation than those far apart, indicating systematic error with respect to sky coordinates.
	\label{fig:dist_vs_match}
}
\end{figure}

\begin{figure}
\centering
\includegraphics[width=.7\textwidth]{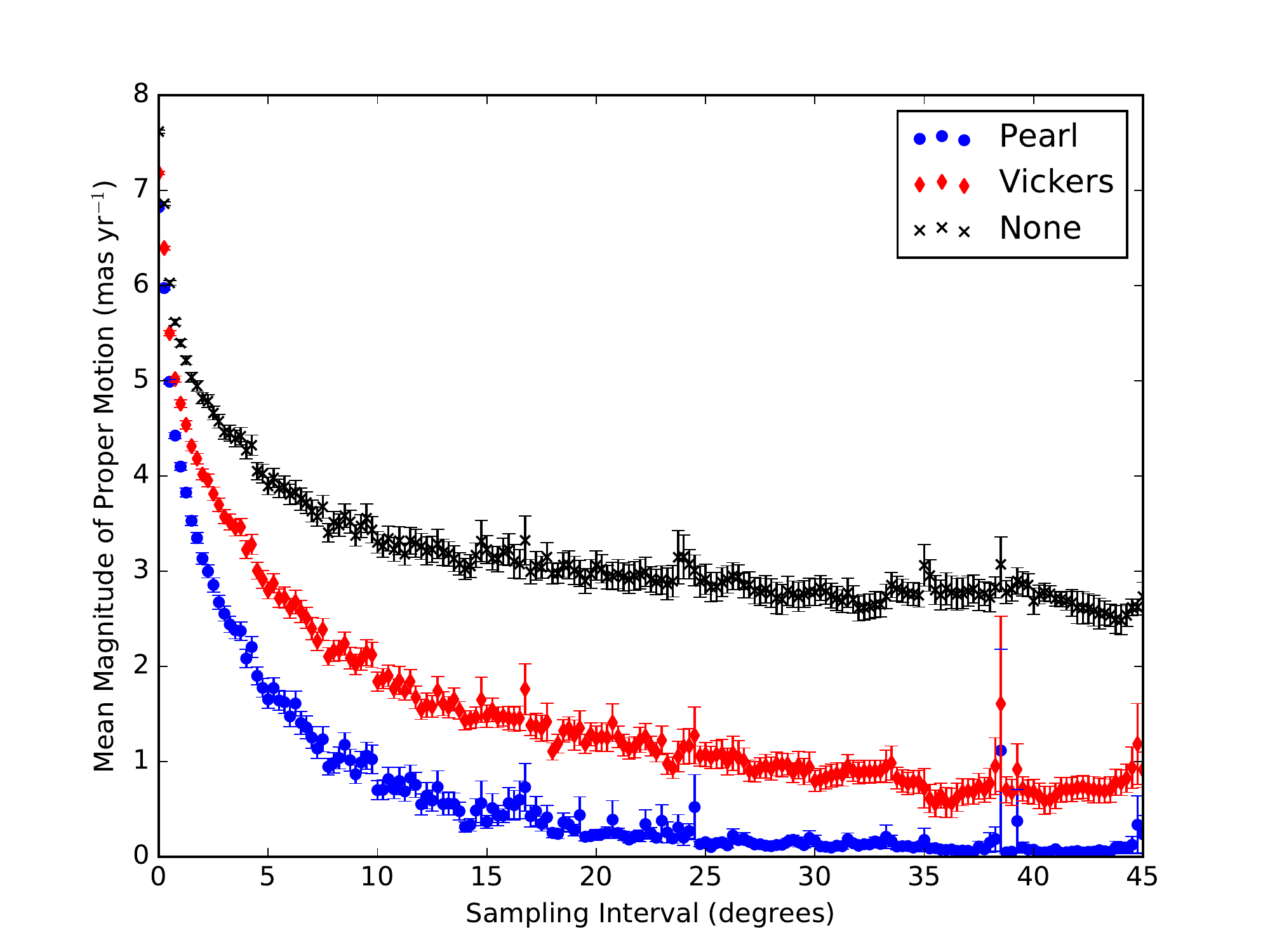}
\caption{
	Average magnitude of proper motion at different sampling rates.
	Higher sampling intervals correspond to larger bins in which proper motions are first averaged together before calculating a magnitude, therefore giving a greater chance of small magnitudes via cancelling out, assuming no lack of systematic error.
	The proper motion distribution demonstrates significantly smaller scale systematic error after applying our [Pearl] correction than those without correction, or even those corrected by the \citet{2016AJ....151...99V} correction.
	\label{fig:subsampling}
}
\end{figure}

Both criteria were tested partially by eye through attempting to eliminate large values and obvious systematics in Figures~\ref{fig:qag_none}~vs.~\ref{fig:qag_vick}~vs.~\ref{fig:qag_pearl}.
Additionally, we measured the second criterion via the trend of the average magnitude of proper motion at various sampling rates (Figure~\ref{fig:subsampling}).
This test gradually increases the subsampling interval, starting at zero degrees, which means the magnitude of the proper motion of each individual object is calculated and averaged.
The interval was tested up to 45$^{\circ}$, at which each 45$^{\circ}\times45^{\circ}$ bin first averages the proper motions of all objects within it and should be close to zero due to cancellation of many positive and negative values, resulting in a near zero average of the magnitude of each bin if there is no net systematic shift.
Since we are comparing the same set of objects in exactly the same bins in Figure~\ref{fig:subsampling}, a trend which approaches zero mean magnitude of proper motion more quickly indicates a smaller amount of systematic error.
In other words, the first value at a sampling rate of zero is predominantly indicative of the magnitude of random error in proper motion.
At larger sampling rates, the value is dominated by systematic error, if present.


\subsection{Estimation of Remaining Systematic Error} \label{sec:calc_se}

Using our correction method, we are also able to produce an estimation of the accuracy of the correction.
Since we subtract the mean proper motion of a set of data to create the correction, the uncertainty is simply the standard error of the mean of this data set.
This uncertainty in the correction is how we calculate our estimation of remaining systematic error, since any statistical error in the correction will propagate through every star in its particular region of the sky.
Therefore, systematic error (in the mean) of each proper motion component is calculated as
\begin{equation}
SE_{\mu} = \frac{\sigma_{\mu}}{\sqrt{n}},
\end{equation}
where $n$ is the number of nearby quasars and $\sigma_{\mu}$ is the standard deviation of their PPMXL proper motion measurements in the respective component.
This value is included along with every correction value in order to determine our confidence in proper motion measurements, and therefore estimate the remaining systematic error present, in each region of the sky.

\subsection{Correction Table} \label{sec:corr_table}

\begin{figure}
\centering
\includegraphics[width=.8\textwidth]{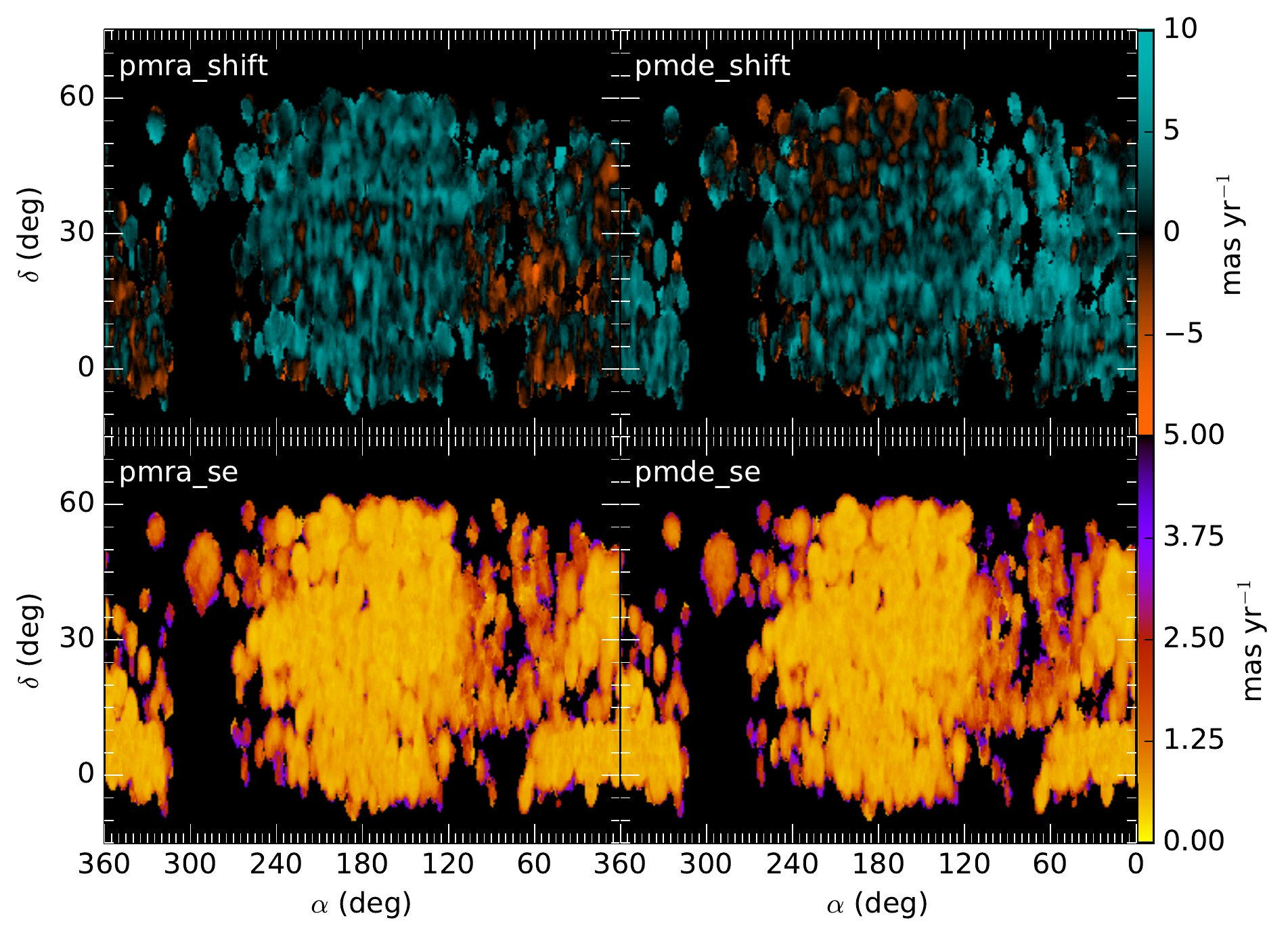}
\caption{
	Columns from our correction table (Table~\ref{tab:pearl_corr}, data file included, \texttt{pearl\_corr.fits}). Each row corresponds to a bin in $(\alpha,\delta)$. 
	The proper motion correction values in each component are given by columns \texttt{pmra\_shift} and \texttt{pmde\_shift}. The uncertainty of these corrections to one standard deviation are given by columns \texttt{pmra\_se} and \texttt{pmde\_se}, respectively. These are used to estimate the remaining systematic error, $SE_\mu$.
	\label{fig:pearl_corr_map}
}
\end{figure}

We constructed a table of bins over very short intervals and spanning the entire region of the sky covered by LAMOST, in which the center position of each bin was assigned a correction value via the method described in Section~\ref{sec:method}.
This resulted in a very long proper motion correction table which is a high resolution representation of the sky between a declination of $-15^{\circ}$ and $75^{\circ}$, where each row corresponds to a bin of size $0.25^{\circ}$ by $0.25^{\circ}$.
Entries in our table with a value of 100.0 in the systematic error columns represent a location lacking sufficient data ($n < 4$) to justify any attempt at correction, and therefore contain a value of zero in their proper motion shift columns.
Note there is only sufficient data in the range of $-10.5^{\circ} < \delta < 61.75^{\circ}$, however, we defined our range of declination using round numbers for more convenient implementation of the correction.
We present our correction table in Table~\ref{tab:pearl_corr}, in order to show the format of each column, and also in Figure~\ref{fig:pearl_corr_map}, which is an image representation of the entire correction and where it is reliable.

The columns of our correction table are as follows:
\verb|"ra"| and \verb|"dec"| give the center values of $\alpha$ and $\delta$ of each bin, in degrees. For a star to be assigned to a bin, it must be within a range of $\pm 0.125$ from the given center values in both $\alpha$ and $\delta$.
\verb|"pmra_shift"| and \verb|"pmde_shift"| give the value which needs to be added to the uncorrected proper motion from PPMXL in each component, in~mas~yr$^{-1}$.
\verb|"pmra_se"| and \verb|"pmde_se"| give the expected remaining systematic error for this location in the sky for their respective components, as calculated in Section~\ref{sec:calc_se}.
Note that proper motion in right ascension is $\mu_{\alpha}\cos{\delta}$, while proper motion in declination is simply $\mu_{\delta}$.


\subsection{Correction Analysis} \label{sec:corr_analysis}

In the vast majority of the sky, the systematic error in proper motion was significantly decreased by our correction, even compared to the current standard correction to the PPMXL database \citep{2016AJ....151...99V}.
This can be seen in both Figure~\ref{fig:qag_pearl} and Figure~\ref{fig:subsampling}.
The difficulty in creating a large area correction to PPMXL seems to be due to the nature in which the systematic error arises.
The error is clumped randomly throughout the sky (Figure~\ref{fig:qag_none}), on the order of only a few degrees, which may suggest that the systematic error is dependent on the photographic plate on which it was measured.
This makes fitting the systematic error to any analytic function extremely difficult, and justifies the use of our correction table (Table~\ref{tab:pearl_corr}, Figure~\ref{fig:pearl_corr_map}).

The area of greatest concern with this as well as many other proper corrections is unfortunately near the Galactic plane.
It appears that our correction still does slightly better at minimizing the systematic error in this region, but it is difficult to compare due to the small amounts of data.
However, our correction may still be more useful in this region because it includes an estimation of remaining systematic error, which is inversely correlated with the number of nearby extragalactic objects from LAMOST, and directly with the dispersion of nearby extragalactic proper motion measurements from PPMXL.
This allows for one to easily identify which parts of the data are meaningful, which is extremely important in discovering actual velocity substructure of disk stars and suppressing deceiving trends caused by systematic error.

\section{Disk Velocity Substructure}
\subsection{Coordinate System and Standard of Rest} \label{sec:coords}

In this paper, we adopt the same Galactic Cartesian $(X,Y,Z)$ and right-handed Galactic cylindrical $(R,\theta,Z)$ coordinate systems used in \citet{2013ApJ...777L...5C} to represent positions. 
Note that both coordinate systems are Galactocentric, and $\theta$ increases in the same direction as Galactic longitude, with the Sun centered at $\theta = 0$.
We use $(V_X,V_Y,V_Z)$ and $(V_R,V_\theta,V_Z)$ to represent the Galactic Standard of Rest velocities in these coordinate systems.
Note that the disk rotates opposite to the $\theta$ direction, with a local velocity of approximately $V_{\theta} = -210$~km~s$^{-1}$.
In these coordinates, we adopt a solar position and velocity of $(X,Y,Z)_{\odot} = (-8,0,0)$~kpc and $(V_X,V_Y,V_Z)_{\odot} = (10.1, 224.0, 6.7)$~km~s$^{-1}$, respectively \citep{2005ApJ...629..268H}.

\subsection{Stellar Selection from LAMOST} \label{sec:lamost}


For this work we use catalogued radial velocities and stellar parameters from the LAMOST survey, including all objects available internally to collaborators through quarter 2 of the fourth observing season (DR4Q2; including the Pilot Survey, Data Release 1: DR1, DR2, and DR3). Our initial stellar sample includes all objects classified as ``STAR'' by the LAMOST stellar parameters pipeline. 
Many stars in the LAMOST catalogs have been observed multiple times over the $\sim5$~years the telescope has operated; for these objects, we combine the velocities, stellar parameters, and distances via a weighted average, with weights determined by the spectroscopic $S/N$. The final catalog after removal (and combination of) duplicate measurements includes $\sim3.1$~million unique stars, of which roughly 3.04~million have $r$-band $S/N$ larger than 10 (median $S/N_r \sim 54$). The median uncertainties in measured/derived spectroscopic quantities for the $S/N_r > 10$ sample are $\sigma_{\rm RV} = 13.5$~km~s$^{-1}$, $\sigma_{T_{\rm eff}} = 114$~K, $\sigma_{\rm [Fe/H]} = 0.15$, $\sigma_{\log{g}} = 0.40$, and $\sigma_{\rm dist} = 0.25$, where $\sigma_{\rm dist}$ is the median of the fractional distance errors, $\frac{|\delta d|}{d}$. This catalog was matched positionally (with a maximum matching radius of $1.5''$) to the PPMXL \citep{2010AJ....139.2440R} catalog of proper motions.

\subsection{Coordinate Transformation} \label{sec:transformation}

We calculated the proper motion of each star in each component ($\mu_\alpha \cos{\delta}$ and $\mu_\delta$) by adding its uncorrected proper motion from PPMXL to the correction from the row in Table~\ref{tab:pearl_corr} corresponding to the star's position.
We also assigned a value of systematic error $SE_{\mu}$ to each component of proper motion, as determined by the corresponding row from Table~\ref{tab:pearl_corr} for each star, in addition to the given statistical random error from the PPMXL catalog.
We derive distances to stars, and their uncertainties, based on the method of \citet{2015AJ....150....4C}, which uses a Bayesian method of estimating the absolute magnitude of each star via comparison of its measured stellar parameters ($T_{\rm eff}$, [Fe/H], and $\log{g}$) to a grid of isochrones.
We use the distance $d$ from an error weighted mean of the distance calculations made from the the estimated absolute magnitude in the 2MASS $K$-band and the SDSS $r$-band, whose extinction-corrected \citep{1998ApJ...500..525S} apparent magnitudes are obtained from LAMOST.
We use radial velocity $v_r$, and its statistical random error, as measured by LAMOST.

We transform the equatorial positions $(d,\alpha,\delta)$ and velocities $(v_r,d\mu_\alpha \cos{\delta},d\mu_\delta)$ of every star into Galactic cylindric coordinates $(R,\theta,Z)$ and $(V_R, V_\theta, V_Z)$.
We also transform and propagate all statistical errors $\sigma$ and systematic errors $SE$ into the same Galactic cylindric coordinates.

\subsection{Selection Cuts}

\begin{figure}
\centering
\includegraphics[width=.8\textwidth]{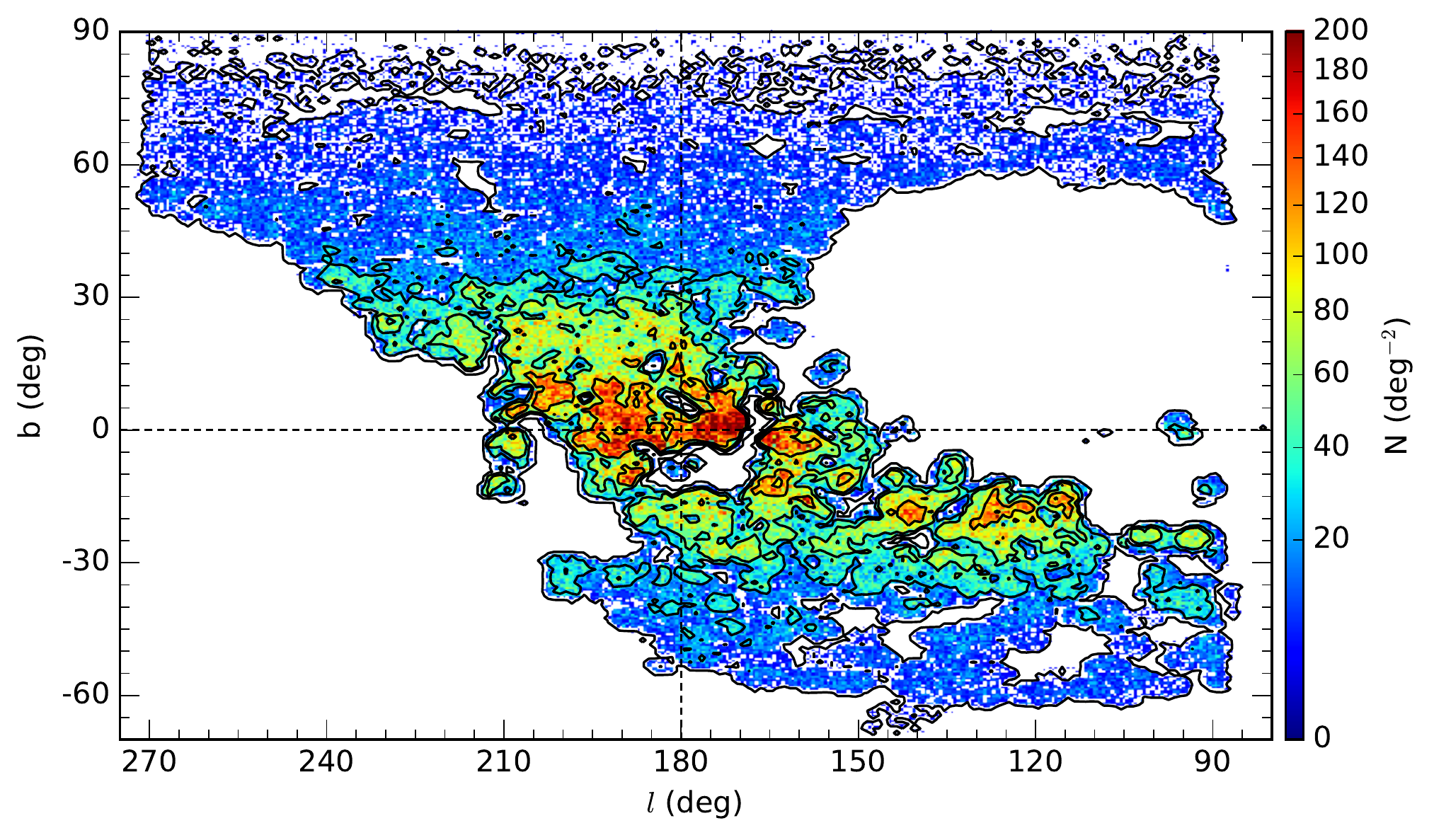}
\caption{
	Density histogram by Galactic longitude and latitude of stellar sample after all selections.
	The abrupt gaps in the data are due to regions with fewer than 4~extragalactic sources from LAMOST within a $2.5^\circ$ radius. We rejected all data in these regions due to their very large systematic uncertainty.
	\label{fig:star_hist_lb}
}
\end{figure}

\begin{figure}
\centering
\includegraphics[width=.8\textwidth]{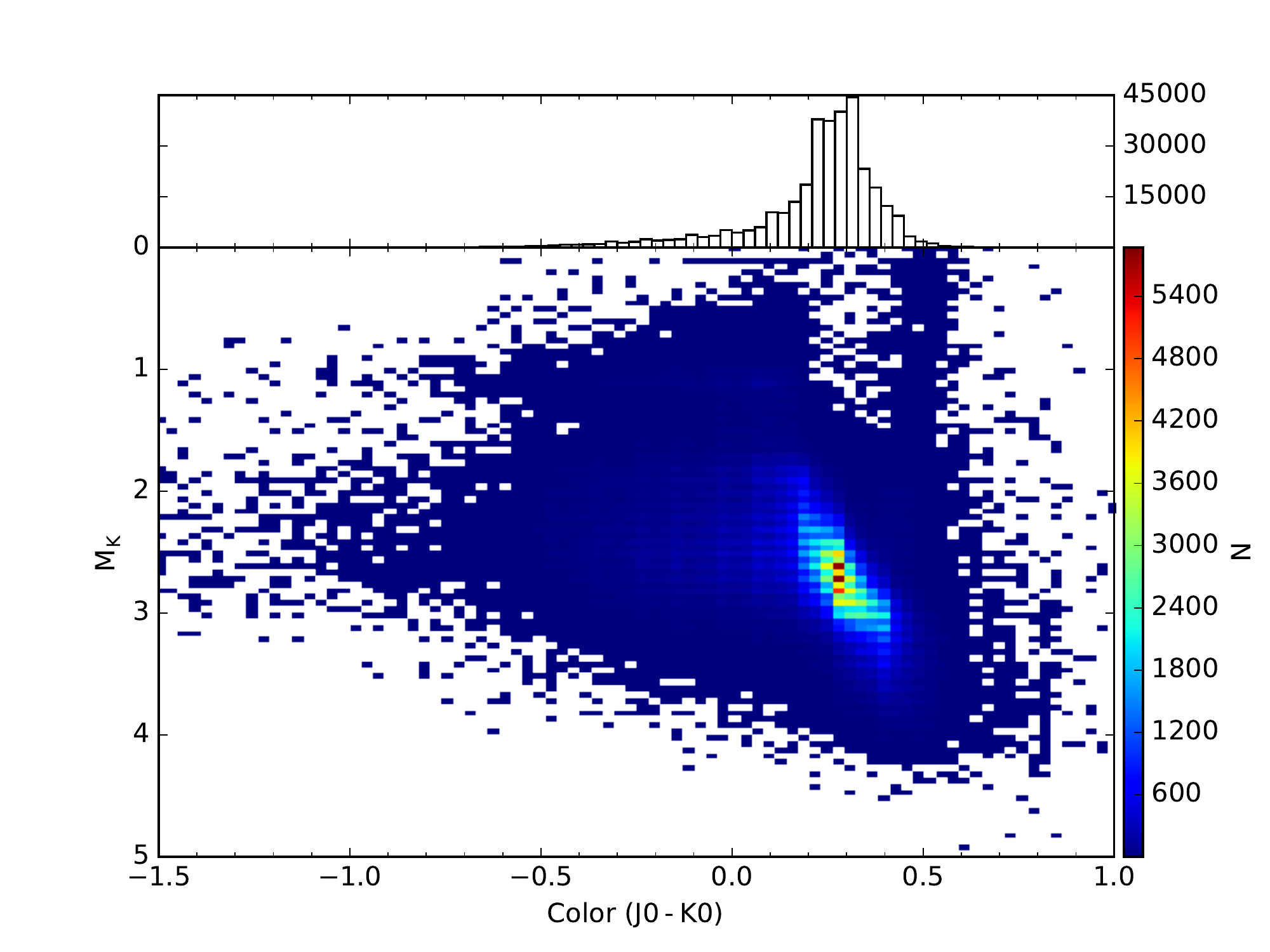}
\caption{
	HR Diagram of our stellar selection.
	Histogram in the top panel displays the $J_0 - K_0$ color distribution of our data, which is centered around~0.3, as is expected of our F stars which were identified in the LAMOST spectroscopic survey.
	\label{fig:hrplot}
}
\end{figure}

We reduced our data set down to 335,599~stars via many selection cuts (a full list of our cuts are shown in Table~\ref{tab:star_cuts}).
In order to conserve direct comparability with the results from \citet{2013ApJ...777L...5C}, we use only F-type stars by selecting stars which have been classified by LAMOST with a subclass beginning with ``F''. 
Our $R$ cut confines the data to a 2~kpc ring beyond the solar radius, while the $\theta$ and $Z$ cuts are intended to be as inclusive as possible; any data outside of this region were far too sparse to be used.
Our velocity cuts are intended to throw away outliers which are most likely not disk stars, but rather nearby high velocity halo stars.
We cut random error and signal to noise ratios $(S/N)$ in order to remove potential outliers, while systematic error cuts remove some especially bad regions of the sky where the systematic error may be very large, due to little or no extragalactic data.
All position and velocity cuts were performed after the coordinate transformation described in Section~\ref{sec:transformation}.
Note that these cuts only define the volume of study, remove statistical outliers and error code values, and remove data with large systematic or random errors.
A heat map of the Galactic coordinates of our stellar sample is shown in Figure~\ref{fig:star_hist_lb}.
We also present an HR Diagram of our stellar sample in Figure~\ref{fig:hrplot}.


\subsection{Spatially Binned Velocity Substructure} \label{sec:binning}

We binned our stellar data into approxiately $0.2 \times 0.2 \times 0.2$~kpc large spatial bins in $R$,$Z$, and $\theta$ (we use $1.3^\circ$ bins in $\theta$, which have a length between 0.18~and~0.23~kpc within $8<R<10$~kpc).  These bin sizes were selected because the distance error for each star is about 20\% \citep{2015AJ....150....4C}, so the positions of stars are typically smoothed out along our line of sight by 200 pc. Because the coordinate system in which we are observing the bulk motions is Galactocentric and the direction of smoothing is heliocentric, there are significant errors in our knowledge of the spatial coordinates of stars in all three axes ($R,\theta,Z$). We did not observe variations in the mean velocities of stars at scales smaller than 0.2 kpc, but that does not mean that smaller scale substructure does not exist; it would have been blurred out by errors in our knowledge of the location of each star.

In each bin, we calculated the number of stars $N$, the mean velocity in each component $(\langle V_R \rangle, \langle V_\theta \rangle, \langle V_Z \rangle)$, the mean systematic error value of each velocity component $(\langle SE_{V_R} \rangle, \langle SE_{V_\theta} \rangle, \langle SE_{V_Z} \rangle)$, and the propagated error in each velocity component ($\sigma_{V_R}, \sigma_{V_\theta}, \sigma_{V_Z}$).
In our analysis, we measured the uncertainty of each velocity component in each bin by summing in quadrature $\langle SE_V \rangle$ and $\sigma_{V}$.

We present these three dimensional mean velocities, along with their uncertainties, in Figure~\ref{fig:vel_sideview} by averaging over the bins in one dimension per panel in order to get the best overall representation of the data from each perspective.
Since the bins that we combined are separated by large distances ($>3^\circ$) in the sky, the systematic errors should be uncorrelated (note that in Section~\ref{sec:corr_analysis} and from Figure~\ref{fig:subsampling} we deduced that any remaining systematic error is only correlated to a few degrees in the sky), so we did not average all systematic error values this time.
Instead, we simply propagate the uncertainties of each bin ($\langle SE_V \rangle$ and $\sigma_{V}$ summed in quadrature), thereby substantially reducing the uncertainty with the number of bins.

In order to show exactly where our data lie, we created Figure~\ref{fig:step_theta} with the individual $0.2 \times 0.2 \times 0.2$~kpc bins by stepping through $1.3^\circ$ slices of $\theta$ in each panel.
It is important to compare apparent trends from Figure~\ref{fig:vel_sideview} to the individual bins in this plot to be sure that the trends are not misleading due to regions of missing data.
Along with Figure~\ref{fig:step_theta}, we present the data file of individual bins containing each component of $\langle V \rangle$, $\sigma_{V}$, and $\langle SE_{V} \rangle$.

Although our data set only encompasses a small portion of the Milky Way disk, it currently contains the most precise bulk velocity calculations to date, as well as estimates of their precision.
This type of data could be very useful in fitting to simulations of the Milky Way disk, and would be appropriate to use with a fitting algorithm that assigns higher weights to data with smaller uncertainties.
We also hope to increase the scope of this data in the future by incorporating the observations from additional facilities in order to cover a larger region of the sky.


\section{Analysis of Disk Substructure}
 
\begin{figure}
\centering
\includegraphics[width=.6\textwidth]{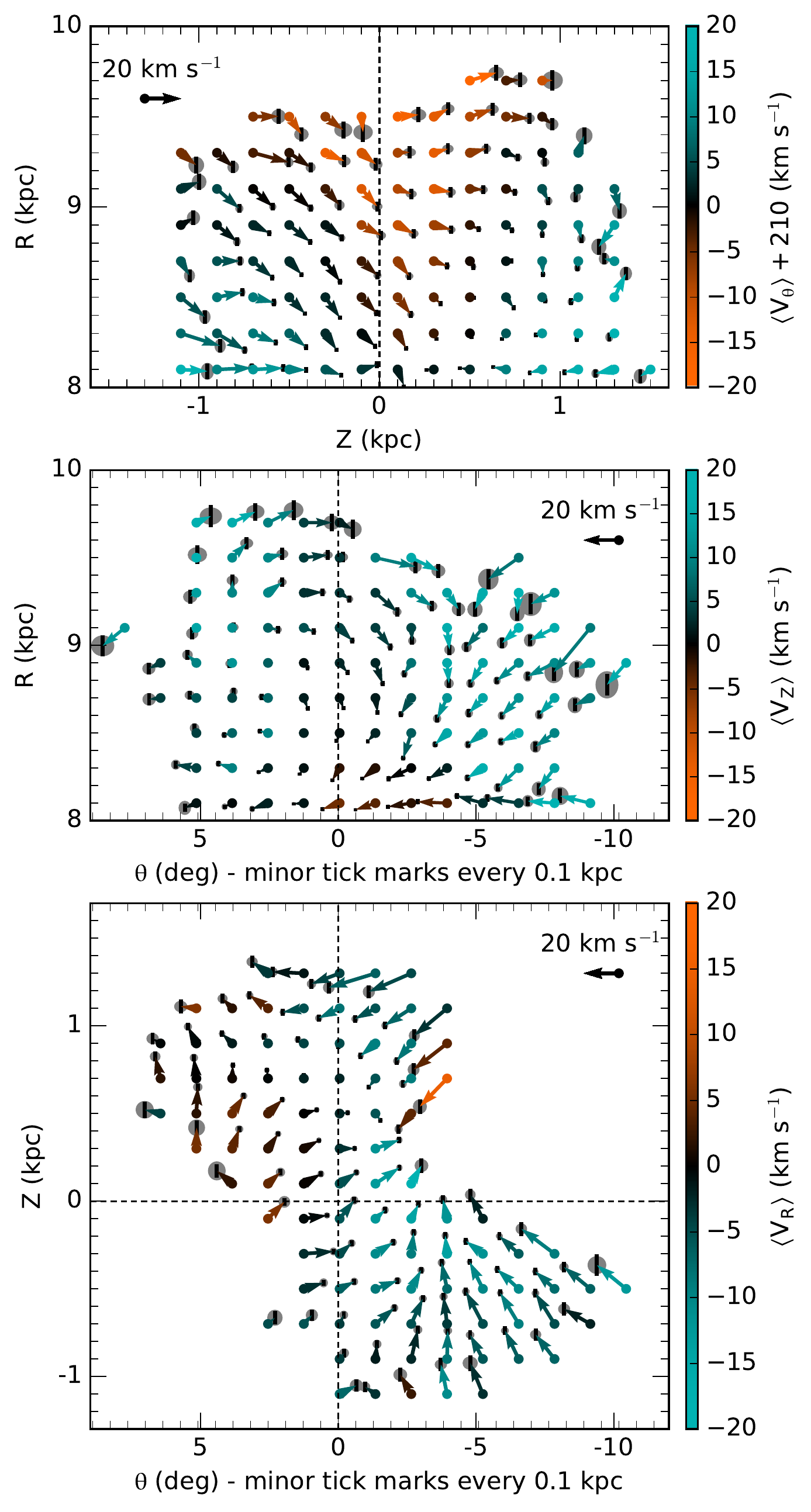}
\caption{
	Several perspectives of stellar velocity averaged over the perpendicular spatial axis.
	Velocity that is coplanar with these axes is measured by the respective components of each vector which corresponds to the bin centered at the tail of the vector.
	The perpendicular velocity is measured by the color of the vector, as scaled by the colorbar.
	The grey ellipses centered around the tip of the arrow measure one standard deviation of the combined uncertainty (statistical and systematic error, added in quadrature) in the two dimensions displayed by the vector, while the half lengths of the black vertical bars measure the combined uncertainty in the color (perpendicular velocity).
	All measures of uncertainty, as well as the planar velocity vectors, follow the displayed scale, which is set such that one minor tick on either axis is equal to 10~km~s$^{-1}$.
	The minor ticks also correspond to approximately 0.1~kpc in the spatial dimensions.
	Each plot is oriented such that blue indicates the velocity is coming out of the page, while orange indicates the velocity is going into the page.
	We only display measurements representing at least fifty data points to present only the most reliable measurements.
	\label{fig:vel_sideview}
}
\end{figure}

The data in Figure~\ref{fig:vel_sideview} are directly comparable with those in Figure~2 of \citet[][hereafter \citetalias{2013ApJ...777L...5C}]{2013ApJ...777L...5C}.  Both analyses use F turnoff stars with radial velocities from LAMOST, and proper motions that have been systematically corrected using extragalactic objects.  The difference is that our current work uses a larger data set available from the ongoing LAMOST survey,  much better distance estimates (with errors) calculated from spectral parameters \citep{2015AJ....150....4C}, and improved systematic corrections (also with errors) to the proper motions.  The addition of error estimates for distances and proper motions has made it possible to both eliminate data with large errors and to demonstrate that the observed bulk disk motions are significant.  In addition, the presentation of the figure has been improved.
 
We first compare the bottom panel of Figure~\ref{fig:vel_sideview} to the bottom left panel of Figure~2 in \citetalias{2013ApJ...777L...5C}.  Both panels show a vertical squeezing mode of the disk in the second quadrant ($\theta<0$). In contrast, in the third quadrant ($\theta>0$), both our data and that of C13 show a small possible expansion mode in the $Z$-direction; however, the only data supporting this mode are north of the Galactic plane, as there is very little data south of the Galactic plane in the third quadrant. Note that the vast majority of the data from LAMOST are located in regions where $V_Z$ is positive: south in the second quadrant (compression mode) and north in the third quadrant (possible expansion mode).
 
A comparison of the lower panel in Figure~\ref{fig:vel_sideview} to the upper left panel of Figure~2 in \citetalias{2013ApJ...777L...5C} also shows substantial agreement in the $V_R$ measurements (note that the color bars are nearly opposite in these two figures).  The velocities are substantially inward towards the Galactic center at $(\theta, Z)=(-2.5^\circ, 0)$, and are nearly zero at $Z=-1$~kpc.  The $V_R$ velocities are nearly zero in the third quadrant ($\theta>0$) in the areas of overlap between our data and that of \citetalias{2013ApJ...777L...5C}.  The area that \citetalias{2013ApJ...777L...5C} finds to have substantially positive $V_R$ at high $Z$ in the second quadrant is not observed in our data due to our elimination of data with larger errors.
 
The top panel of Figure~\ref{fig:vel_sideview} can be directly compared with the right panels of Figure~2 in \citetalias{2013ApJ...777L...5C}, and again shows significant agreement. The stars below the Galactic plane are streaming towards the Galactic plane and towards the Galactic center. However, we do not observe the same region north of the Galactic plane moving down towards the plane as they are in \citetalias{2013ApJ...777L...5C}. We only observe squeezing in the north at close $R$, as opposed to the increasing trend of squeezing with $R$ according to \citetalias{2013ApJ...777L...5C}. However, this apparent lack of downward motion in the north is largely due to the overwhelming amount of data in the northern third quadrant, which are at rest or slightly expanding, compared to the few stars in the northern second quadrant, which are squeezing (this is clear from Figure~\ref{fig:step_theta}). As in \citetalias{2013ApJ...777L...5C}, we see the stars moving towards the Galactic center just south of the Galactic plane, but staying nearly stationary in Galactocentric radii everywhere else between $-1<Z<1$~kpc.
 
\citetalias{2013ApJ...777L...5C} did not explore the circulation velocity of stars ($V_\theta$).  We show in the top panel of Figure~\ref{fig:vel_sideview} that the circulation velocity is larger near the Galactic plane than it is at distances of 1~kpc above and below the plane at the position of the Sun.  Moreover, we find that the average circulation velocity is anywhere from 5~to~20~km~s$^{-1}$ larger (more negative in our coordinate system) 1.5~kpc further from the Galactic center than it is at the position of the Sun (Figures~\ref{fig:rotationcurve}). 

In order to better quantify the rotation curve, we averaged together the angular velocities of the bins in the thin disk ($Z \le 0.2$ kpc) and the bins representing the thick disk ($Z > 0.2$ kpc), and show the results in Figure~\ref{fig:rotationcurve}.  The thin disk rotation velocity increases from 207~km~s$^{-1}$ at 0.1~kpc outside the solar circle to 225~km~s$^{-1}$ 1.5~kpc outside the solar circle.  Over the same Galactocentric radii, the thick disk rotation velocity increases from 196~km~s$^{-1}$ to 218~km~s$^{-1}$.  The increasing trend in disk speed at greater radii is comparable to the trend found in \citet{2014A&A...563A.128L}.

 
The middle panel of Figure~\ref{fig:vel_sideview} looks down on the disk (with a slight distortion because the radial lines of constant $\theta$  from the Galactic center are vertical and parallel in this diagram).  It shows little motion in the third quadrant, outside the solar circle.  There is a large streaming in the second quadrant up (positive $Z$), inward, and in a circulating pattern in $(R, \theta)$.  Note that our sampling of stars in the second quadrant is dominated by stars south of the Galactic plane.  In order to remove doubt that this circulation is caused by an inconsistency in the availability of data at different heights above the Galactic plane, we show the same data for only stars within 0.2~kpc of the Galactic plane in Figure~\ref{fig:thindisk}.  This streaming motion is real, and present at very low Galactic latitudes.

It is tempting to relate the observed streaming motion to the spiral structure of the Milky Way, but the relationship is not clear.  For example, a series of papers by \citet{2014MNRAS.443.2757K}, \citet{2015MNRAS.450.2132H}, and \citet{2015MNRAS.453.1867G} explore the density and velocity substructure of disks containing spiral structure in N-body/smoothed particle hydrodynamics simulations of Milky Way-like galaxies.  They conclude that on the trailing side of a spiral arm, disk stars move outwards from the Galactic center with slower rotation speed; and on the leading side of the spiral arm they move inwards towards the Galactic center with faster rotation speed \citep{2014MNRAS.439..623G}.  This predicts a shearing in velocity substructure which might or might not be related to the substructure we identify.
 
In Figure~\ref{fig:step_theta}, we break down the top panel of Figure~\ref{fig:vel_sideview} into narrow slices of $\theta$, so that effects within our three dimensional data cube are not averaged over.  Each $1.3^\circ$ angular slice around $R\sim9$~kpc is about  0.2~kpc thick (calculated from $9 \tan{1.3^\circ}$).  It is clear from Figure~\ref{fig:step_theta} that bulk velocities shift measurably on scales of 0.2~kpc, in all three spatial dimensions.  Since we averaged over about 2~kpc in each dimension in Figure~\ref{fig:vel_sideview}, and there were substantial portions of the $0.2 \times 0.2 \times 0.2$~kpc region in which data were missing, the figure must be interpreted with some care.  The conclusions drawn from Figure~\ref{fig:step_theta} represent a more accurate picture of the velocity substructure.
 
Figure~\ref{fig:step_theta} shows that the velocities in some regions of the disk differ from nearby regions by 20~km~s$^{-1}$ or more.  Therefore, measurements of $V_\theta$, or any other bulk measurement of the disk, will depend on the region sampled.  At various places in the $(X,Y)$ plane (represented in our diagrams as $(R, \theta)$), within $|Z|<0.2$~kpc, we measure coherent circulation velocities below 210~km~s$^{-1}$, above 230~km~s$^{-1}$, and everywhere in between (Figure~\ref{fig:thindisk}).  This underscores the need to carefully define what one means by the circulation velocity of the Milky Way, or even the local standard of rest, and to carefully define the sample used to measure it. 

 
The right column of panels in Figure~\ref{fig:step_theta} show a squeezing mode of the disk.  In the panels with $-0.7^\circ>\theta>-3.3^\circ$, the stars are squeezing towards $Z=0.5$~kpc at $R=9$~kpc, and towards $Z=0.2$~kpc at $R=8$~kpc.  This could signify that we are seeing the results of both bending and breathing modes of the disk.  It is interesting to note that the midplane of the disk is thought to move 70~pc towards positive $Z$ at 2~kpc from the Sun \citep[corresponding to our $R=10$~kpc;][]{2015ApJ...801..105X}, which is a much smaller shift than the apparent center of the (compression) breathing mode observed in our data.
 
The measured Galactocentric radial velocities are mostly small, or fraught with large errors.  However, there are very significant inward velocities of stars of about 15~km~s$^{-1}$ between zero and 1~kpc from the Sun, within five degrees of the anticenter in the second quadrant.  These inward velocities are not observed on the other side of the anticenter, only $5^\circ$ (0.74~kpc) away.

\section{Conclusion}

\begin{figure}
\centering
\includegraphics[width=.9\textwidth]{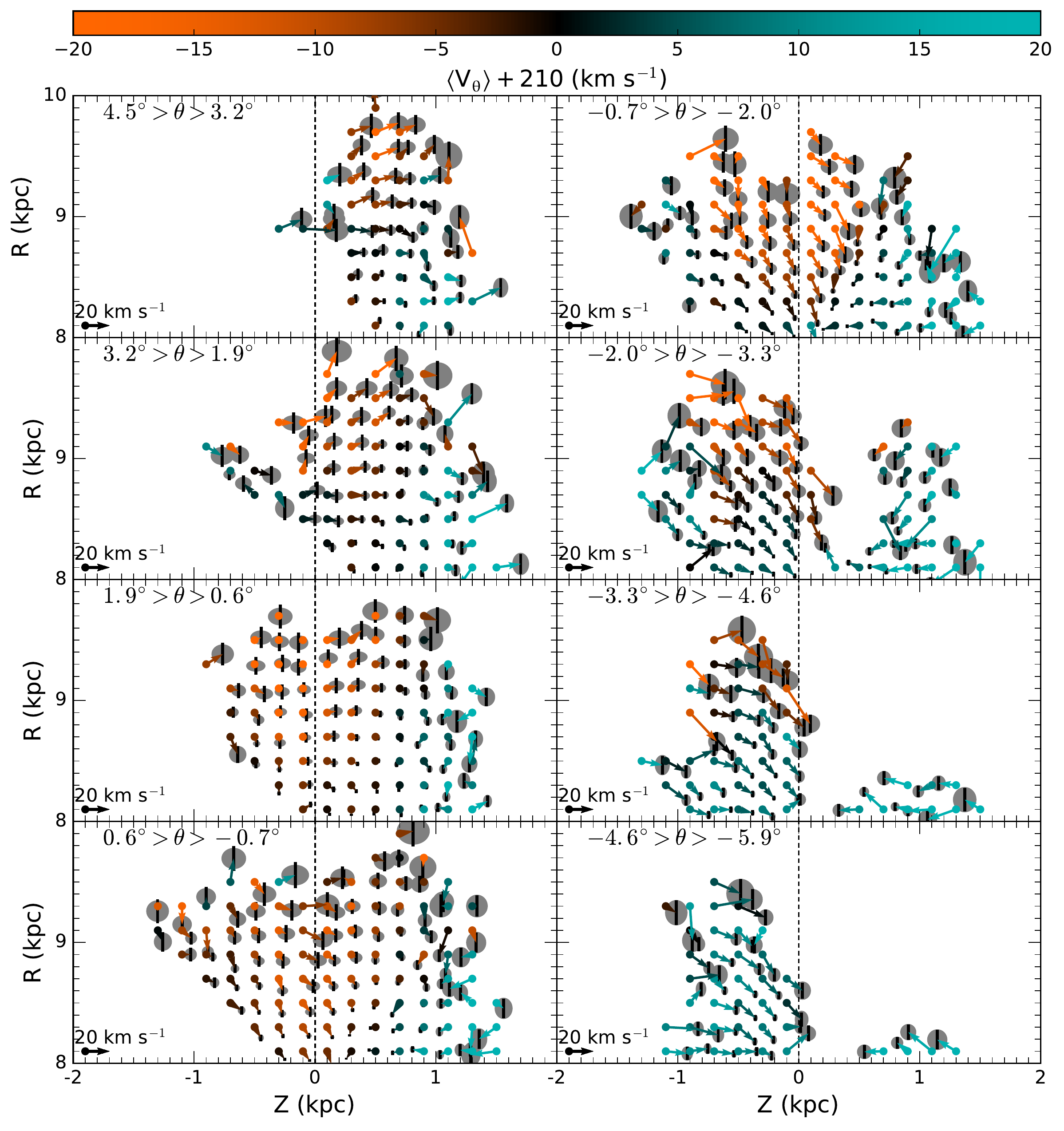}
\caption{
	Stellar velocity with slices in $\theta$ (data file included, \texttt{spatial-bins.fits} or \texttt{spatial-bins.csv}).
	These plots show the individual bins that are within the specified theta bounds.
	Starting at the top left, the plots start in the third quadrant (positive theta) and consecutively step in the negative direction, passing through the anticenter and into the second quadrant.
	We only display bins with at least ten data points in order to eliminate only the measurements with very large uncertainties.
	\label{fig:step_theta}
}
\end{figure}

\begin{figure}
\centering
\includegraphics[width=.8\textwidth]{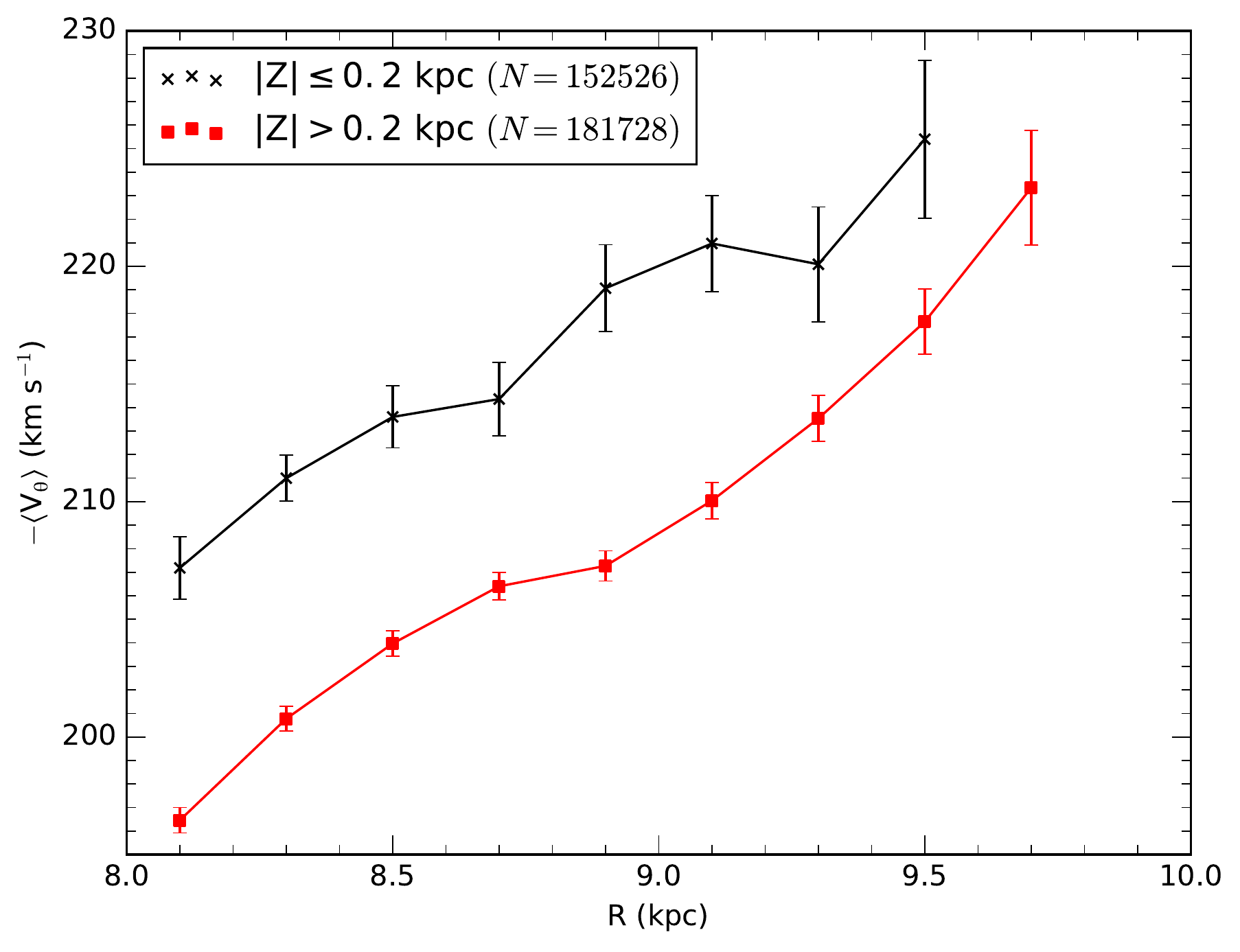}
\caption{
   Rotation curve in the thin and thick disks. The vertical axis measures the mean circulation velocity (which is the opposite direction from $V_\theta$). As in the previous plots, uncertainties are measured by summing in quadrature the standard error of the mean velocity and the average systematic error in each bin. We only display measurements representing at least fifty data points. 
   \label{fig:rotationcurve}
}
\end{figure}

\begin{figure}
\centering
\includegraphics[width=.8\textwidth]{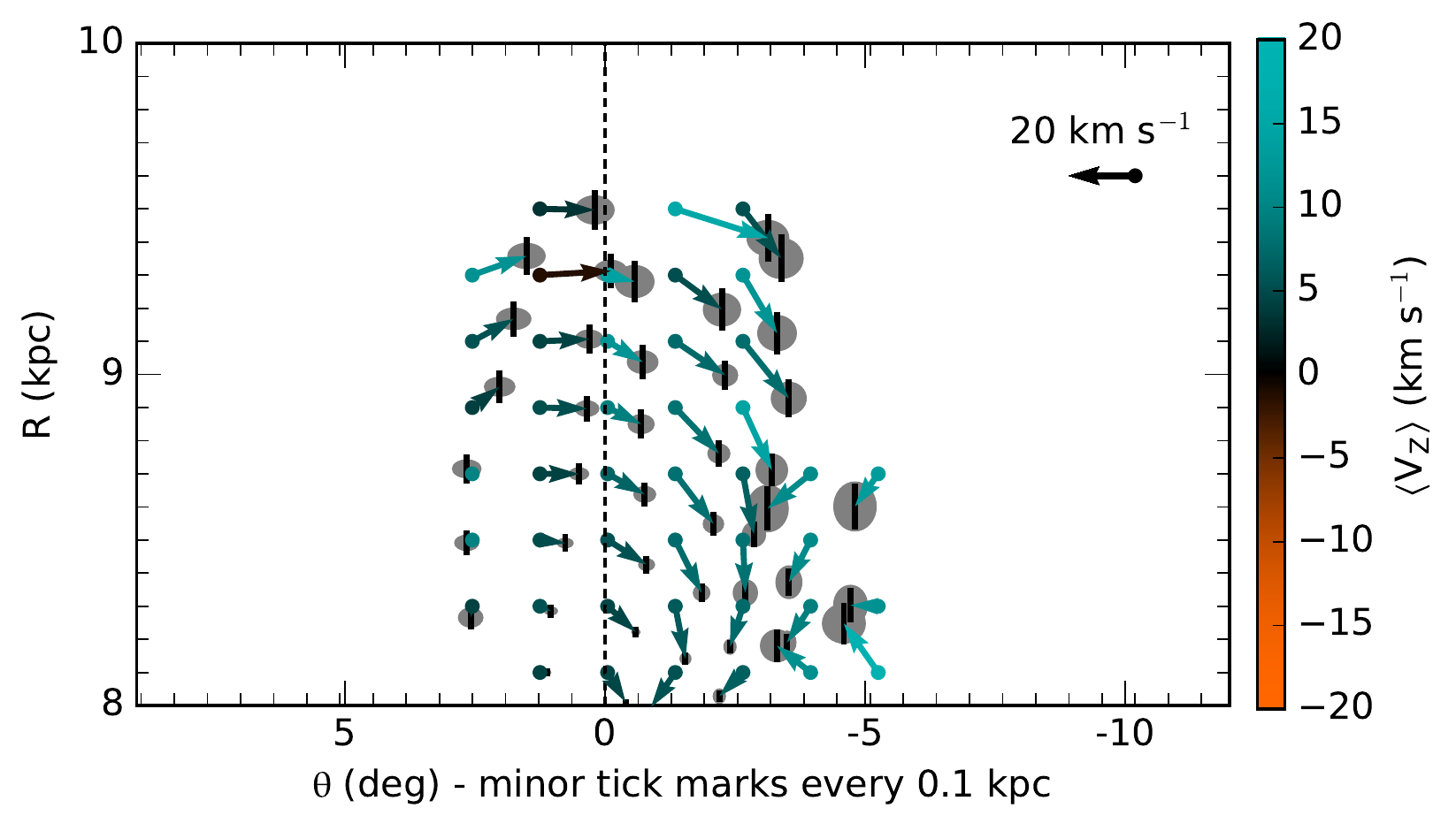}
\caption{
   Stellar velocity in the thin disk. Only data within $|Z| < 0.2$~kpc are included (i.e. only the bins centered at $Z = \pm0.1$~kpc). This amounts to $N\sim150,000$ stars, which is nearly half of the stellar data. Measurements are averaged over bins along the $Z$-axis, as was done in the middle panel of Figure~\ref{fig:vel_sideview}. We only display measurements representing at least fifty data points.
   \label{fig:thindisk}
}
\end{figure}

In this paper, we present the bulk velocities of stars in the Milky Way disk, within about 1.5~kpc of the Sun  and towards the Galactic anticenter.  We use radial velocities and distances of F~stars from LAMOST.   Corresponding proper motions are derived from PPMXL, and have been corrected for systematic errors using a novel technique that is sensitive on angular scales of only a few degrees, which is comparable to the scale of the photographic plates from which many of the proper motion measurements were made.  The disk bulk velocities are given with a spatial granularity of 0.2~kpc, and are available as companion data to Figure~\ref{fig:step_theta}.  A comprehensive error analysis has been performed that includes contributions from distances, proper motions, and radial velocities.  The errors give us confidence that we are measuring  velocity substructure as a function of position in the Milky Way disk.  The results are generally in agreement with the results of \citet{2013ApJ...777L...5C}, but are analyzed in all three dimensions and include error bars that convincingly show that the results are significant.

We show that the bulk velocity of the disk shifts measurably on length scales at least as small as 0.2~kpc in all three dimensions.  The velocity can shift by 20~km~s$^{-1}$ or more over distances of half a kpc.  In particular, the circulation velocity within 0.2~kpc of the Galactic plane shifts from 205~km~s$^{-1}$ near the Sun's position to around 225~km~s$^{-1}$ 1.5~kpc farther away from the Galactic center.  While the observed stars in the third quadrant (within 1~kpc from the Sun, in the direction of the anticenter) are generally moving coherently, at about 210~km~s$^{-1}$ on circular orbits, the stars in the second quadrant (only about a kpc away) exhibit a bulk motion towards the Galactic center of about 15~km~s$^{-1}$ and are also compressing in the $Z$ direction towards a point about 0.5~kpc above the Galactic plane.  Looking down on the disk, the stars appear to move in a circular streaming motion with a radius of several kpc.
As expected, we observe that the magnitude of the circulation velocity around the Galactic center decreases as a function of distance from the Galactic plane.

We provide a correction table for PPMXL proper motions, in the regions surveyed by LAMOST, that can be used by other authors.  We show that the remaining systematic error using our technique is much less significant than that obtained using the \citet{2016AJ....151...99V} correction in most regions of the sky.  In addition, we provide estimates of the remaining systematic error, which allows for easy determination of the reliability of the measurements.

While previous authors have shown that there is spatial and velocity substructure in the Milky Way disk, this paper begins the process of mapping out that substructure.  We plan to continue this work in the future by including additional data from LAMOST as they are obtained, including stars of other spectral types, and adding additional stars from other surveys, including RAVE and Gaia.  Gaia in particular will provide us with much more accurate proper motions.  The velocity substructure maps will ultimately be combined with simulations to explore the spiral structure of our galaxy, and potentially the properties of the dwarf galaxies and subhalos that orbit the Milky Way and are believed to produce the disk ripples that we observe.

\acknowledgements

This project was funded by NSF grants AST 14-09421 and AST 16-15688.
This research made use of matplotlib, a Python library for publication quality graphics \citep{2007CSE.....9...90H}.
This work is based in part on services provided by the GAVO Data Center and the data products from the PPMXL database of \citet{2010AJ....139.2440R}.
This work makes use of the Guoshoujing Telescope (the Large Sky Area Multi-Object Fiber Spectroscopic Telescope LAMOST, \citealt{2012RAA....12.1197C}) which is a National Major Scientific Project built by the Chinese Academy of Sciences. Funding for the project has been provided by the National Development and Reform Commission. LAMOST is operated and managed by the National Astronomical Observatories, Chinese Academy of Sciences.

\facility{LAMOST \citep{2012RAA....12.1197C}}
\software{Python, NumPy, matplotlib \citep{2007CSE.....9...90H}, PyFITS}


\begin{deluxetable}{l}
\tabletypesize{\small}
\tablecolumns{1}
\tablecaption{
	\label{tab:star_cuts}
}
\tablehead{\colhead{Stellar Sample Selection Criteria}}

\startdata
	Classified as F-type star by LAMOST\\
	$S/N > 5$ in g, r, and i\\
	$8 < R < 10$ kpc\\
	$-2 < Z < 2$ kpc\\
	$9^{\circ} > \theta > -12^{\circ}$\\
	$-150 < V_{R} < 150$ km s$^{-1}$\\
	$-400 < V_{\theta} < -100$ km s$^{-1}$\\
	$-150 < V_{Z} < 150$ km s$^{-1}$\\
	$SE_{V_{R}} < 8$ km s$^{-1}$ (Systematic Error)\\
	$SE_{V_{Z}} < 10$ km s$^{-1}$\\
	$SE_{V_{\theta}} < 10$ km s$^{-1}$\\
	$\sigma_{V_{R,Z,\theta}} < 40$ km s$^{-1}$ (Random Error)\\
	$\sigma_{R} < 0.2$ kpc\\
	$\sigma_{Z} < 0.2$ kpc\\
	$\sigma_{\theta} < 1.3^{\circ}$\\
\enddata
\end{deluxetable}

\begin{deluxetable}{ccccccc}
\tablecolumns{7}
\tabletypesize{\footnotesize}
\tablecaption{
	Sample of the data from our proper motion correction table (Figure~\ref{fig:pearl_corr_map}, data file included, \texttt{pearl\_corr.fits}).
	Each of the $(1440 \times 360)$ 518,400~rows corresponds to a bin representing a small region in the sky at which the following correction values apply.
	Each bin is $0.25^\circ \times 0.25^\circ$, spanning $0^{\circ} \leq \alpha < 360^{\circ}$ and $-15^{\circ} \leq \delta < 75^{\circ}$.
	Note that the table only contains useful data between the declinations of $-10.50^{\circ} < \delta < 61.75^{\circ}$, resulting in no correction values given to many rows at the beginning and end of the table; as is the case for the first three displayed rows.
	Rows in which no correction has been attempted are marked with a very large systematic error value of 100.0 in each component.
	\label{tab:pearl_corr}
}
\tablehead{\colhead{Index} & \colhead{\texttt{ra}} & \colhead{\texttt{dec}} & \colhead{\texttt{pmra\_shift}} & \colhead{\texttt{pmde\_shift}} & \colhead{\texttt{pmra\_se}} & \colhead{\texttt{pmde\_se}}}

\startdata
	0 & 0.125 & -14.875 & 0.0 & 0.0 & 100.0 & 100.0 \\
	1 & 0.375 & -14.875 & 0.0 & 0.0 & 100.0 & 100.0 \\
	2 & 0.625 & -14.875 & 0.0 & 0.0 & 100.0 & 100.0 \\
	...    & ...     & ...    & ...   & ...   & ...   & ...   \\
	250000 & 220.125 & 28.375 & 0.359 & 1.032 & 0.487 & 0.610 \\
	250001 & 220.375 & 28.375 & -0.299 & 1.045 & 0.518 & 0.613 \\
	250002 & 220.625 & 28.375 & -0.557 & 0.952 & 0.503 & 0.640 \\
	250003 & 220.875 & 28.375 & -0.577 & 1.234 & 0.495 & 0.649 \\
	250004 & 221.125 & 28.375 & -0.548 & 1.141 & 0.490 & 0.604 \\
	...    & ...     & ...    & ...   & ...   & ...   & ...   \\
\enddata
\end{deluxetable}


%
%
%
%

\bibliographystyle{apj}

\end{document}